\documentclass[aps,english,preprint,nofootinbib]{revtex4}

\usepackage{graphicx}
\usepackage{hyperref}
\usepackage{amsmath, amssymb}
\usepackage{babel}
\usepackage{color}
\usepackage{mathrsfs}
\usepackage{slashed}

\begin{document}

\title{Complete theory of radiative corrections to $K_{\ell 3}$ decays and the $V_{us}$ update}

\author{Chien-Yeah Seng$^{1}$}
\author{Daniel Galviz$^{1,2}$}
\author{Mikhail Gorchtein$^{3,4}$}
\author{Ulf-G. Mei{\ss}ner$^{1,5,6}$}

\affiliation{$^{1}$Helmholtz-Institut f\"{u}r Strahlen- und Kernphysik and Bethe Center for
  Theoretical Physics,\\ Universit\"{a}t Bonn, 53115 Bonn, Germany}
\affiliation{$^{2}$Yau Mathematical Sciences Center,\\ Tsinghua University, Beijing 100084, China}
\affiliation{$^{3}$Institut f\"{u}r Kernphysik, Johannes Gutenberg-Universit\"{a}t,\\ J.J. Becher-Weg 45, 55128 Mainz, Germany}
\affiliation{$^{4}$PRISMA Cluster of Excellence, Johannes Gutenberg-Universit\"{a}t, Mainz, Germany}
\affiliation{$^{5}$Institute for Advanced Simulation, Institut f\"ur Kernphysik and J\"ulich Center
  for Hadron Physics, Forschungszentrum J\"ulich, 52425 J\"ulich, Germany}
\affiliation{$^{6}$Tbilisi State  University,  0186 Tbilisi, Georgia}

\date{\today}

\begin{abstract}

We fill up the missing piece in our own re-analysis of the long-distance electromagnetic radiative corrections to semileptonic kaon decays by performing a rigorous study in the $K\rightarrow\pi\mu^+\nu_\mu(\gamma)$ channels. With appropriate experimental and lattice inputs, we achieve a precision level of $10^{-4}$ in these channels. This is comparable to our previous analysis in the $K\rightarrow\pi e^+\nu_e(\gamma)$ channels. With this new result, we present an updated global analysis to extract the Standard Model parameter $|V_{us}|$ from semileptonic kaon decays. We obtain $|V_{us}|=0.22308(55)$ and $0.22356(73)$, using the lattice average of the $K^0\rightarrow\pi^-$ transition form factor at $N_f=2+1+1$ and $N_f=2+1$,  respectively.

\end{abstract}

\maketitle


\section{Introduction}

A series of new applications of the classical Sirlin representation of the Standard Model (SM) radiative corrections (RCs)~\cite{Sirlin:1977sv,Seng:2021syx} has recently brought our understanding of semileptonic decays of mesons, nucleons and nuclei to the next level. For instance, the implementation of a dispersion relation analysis on top of this framework has significantly improved the accuracy of the SM predictions of the free neutron~\cite{Seng:2018yzq,Seng:2018qru,Seng:2020wjq,Shiells:2020fqp,Gorchtein:2021fce} and nuclear beta decays~\cite{Seng:2018qru,Gorchtein:2018fxl}, leading to a more precise extraction of the top-row Cabibbo-Kobayashi-Maskawa (CKM) matrix element $V_{ud}$. These new developments unveiled an apparent violation of the top-row CKM unitarity~\cite{Zyla:2020zbs}, which provides one of the most promising evidences of the breakdown of precise SM theory predictions at low-energy precision experiments, alongside with the anomalies observed in the muon anomalous magnetic moment~\cite{Fermigm2,Aoyama:2020ynm,Miller:2007kk,Miller:2012opa,Jegerlehner:2009ry} and decays of B-mesons~\cite{Aaij:2019wad,Aaij:2014ora,Aaij:2015yra,Aaij:2015oid}.

It was first thought that Sirlin's representation is only useful for decay processes where the parent and daughter particles are nearly degenerate, because some of the more complicated hadronic matrix elements involving products of three electroweak currents vanish in this limit. However, some of the authors in this paper pointed out that this framework is equally applicable to semileptonic decay with large mass gaps, e.g. the kaon semileptonic decays ($K_{\ell 3}$), upon combining it with more modern techniques such as Chiral Perturbation Theory (ChPT)~\cite{Seng:2019lxf}. In a follow-up work~\cite{Seng:2020jtz} we outlined an efficient approach for lattice Quantum Chromodynamics (QCD) to help fixing the large theory uncertainties in the $K_{\ell 3}$ RC that originate from non-perturbative QCD, which was then partially put into practice~\cite{Ma:2021azh}. With these new theory basis and lattice inputs, we improved the existing theory precision of the $K_{e3}$ RC~\cite{Cirigliano:2008wn} by almost an order of magnitude~\cite{Seng:2021boy,Seng:2021wcf}, and presented an updated global analysis to the value of the top-row CKM matrix element $|V_{us}|$ extracted from semileptonic kaon decays~\cite{Seng:2021nar}. This new analysis supports the current top-row CKM unitarity violation, and further sharpens a previously-observed discrepancy between the values of $|V_{us}|$ obtained from $K_{\ell 3}$ and the leptonic kaon decay ($K_{\mu 2}$)~\cite{Zyla:2020zbs}.

This paper is a generalization of the aforementioned works to cover all channels in semileptonic kaon decays, in particular the $K_{\mu 3}$ channels.
A reason why only the $K_{e3}$ decays were previously focused on is that in these channels the effect of the (more) poorly-constrained RCs to the intrinsically non-forward charged weak form factor $f_-(t)$ to the decay rate is suppressed by $m_e^2/M_K^2\sim 10^{-6}$ (on top of $\alpha/\pi$), making them theoretically clean. In this work we show that, with the precision goal of $10^{-4}$, this can even be done for the $K_{\mu 3}$ channels by first separating the infrared (IR)-divergent and IR-finite corrections to $f_-(t)$, calculating the former precisely and matching the latter to a fixed-order ChPT expression. This simple prescription enables us to calculate the $K_{\mu 3}$ RCs to a same level of precision with their $K_{e3}$ counterparts. 

The content of this work is arranged as follows. In Section~\ref{sec:basic} we introduce the basic notations and formula for the $K_{\ell 3}$ decay rate. In Section~\ref{sec:virtual} we outline the theory framework of the $\mathcal{O}(\alpha)$ virtual corrections, most of them are similar to Refs.~\cite{Seng:2021boy,Seng:2021wcf} except a novel and more rigorous treatment of the RCs to $f_-(t)$. In Section~\ref{sec:real} we discuss the bremsstrahlung contributions. The numerical results and error analysis are presented in Section~\ref{sec:num} together with a new global analysis of $|V_{us}|$ from semileptonic kaon decays. In Section~\ref{sec:concl} we state our final conclusions. 

\section{\label{sec:basic}Basic formalism}

This work is a straightforward extension of our previous $K_{e3}$ calculation. It is rather meaningless to copy everything from the existing literature, so most of the time we simply refer the reader to the notations, formulas and descriptions in specific sections of Ref.~\cite{Seng:2021wcf}. Some trivial generalizations of notations, e.g. $e\rightarrow \ell$, are automatically implied if not stated otherwise.   

First of all, we define $K_{\ell 3}$ (where $\ell=e,\mu$) as the fully-inclusive semileptonic decay process $K(p)\rightarrow \pi(p')+\ell^+(p_\ell)+\nu_\ell (p_\nu)+n\gamma$, with $n\geq 0$. At tree-level, the decay amplitude is given by:
\begin{equation}
M_0=-\frac{G_F}{\sqrt{2}}\bar{u}_\nu \gamma^\mu(1-\gamma_5)v_\ell F_\mu(p',p)~,
\end{equation} 
where 
\begin{equation}
F_\mu(p',p)=\langle \pi(p')|(J_\lambda^W)^\dagger|K(p)\rangle=V_{us}^*[f_+^{K\pi}(t)(p+p')_\mu+f_-^{K\pi}(t)(p-p')_\mu]
\end{equation}
defines the (real) charged weak form factors $f_\pm^{K\pi}(t)$ (for notational simplicity, from now on we will suppress the superscript $K\pi$ whenever it does not cause any confusion). The standard Mandelstam variables are defined as $s=(p'+p_\ell)^2$, $t=(p-p')^2$ and $u=(p-p_\ell)^2$, but for $n=0$ only two kinematic variables are independent, which are  often chosen as $y=2p\cdot p_\ell/M_K^2$ and $z=2p\cdot p'/M_K^2$. The total tree-level decay rate is then given by:
\begin{equation}
(\Gamma_{K_{\ell 3}})_\mathrm{tree}=\frac{M_K}{256\pi^3}\int_{\mathcal{D}_3}dydz \:G_F^2|V_{us}|^2\left\{f_+^2H(+1,+1)+2f_+f_-H(+1,-1)+f_-^2H(-1,-1)\right\}~,
\end{equation}
where the integration region $\mathcal{D}_3$ is defined in Appendix~A of Ref.~\cite{Seng:2021wcf}, while the function $H(a,b)$ is defined in Eq.~(2.7) of the same reference. 

The full $\mathcal{O}(\alpha)$ electroweak RC to $K_{\ell 3}$ include one-loop virtual corrections at $n=0$ and tree-level real (i.e. bremsstrahlung) corrections at $n=1$. It is most conveniently expressed as a fractional correction to the tree-level decay rate:
\begin{equation}
\delta_{K_{\ell 3}}=\frac{\delta \Gamma_{K_{\ell 3}}}{(\Gamma_{K_{\ell 3}})_\mathrm{tree}}~.
\end{equation}
However, one must be aware that in the existing ChPT treatment some of the short-distance electromagnetic corrections are not included in the definition of $\delta_{K_{\ell 3}}$, but rather redistributed into the charged weak form factors as well as the isospin-breaking corrections (see Sec.~8 of Ref.~\cite{Seng:2021wcf} for a discussion). After removing those terms, our result can be directly compared to the ``long-distance'' electromagnetic RC $\delta_\mathrm{EM}^{K\ell}$ studied in the standard ChPT literature by further removing a short-distance, channel-independent electroweak correction factor:
\begin{equation}
\delta_\mathrm{EM}^{K\ell}=\delta_{K_{\ell 3}}-(S_\mathrm{EW}-1)~,
\end{equation}
where $S_\mathrm{EW}=1.0232(3)_\mathrm{HO}$~\cite{Marciano:1993sh}.

Below we discuss the virtual and real corrections separately.

\section{\label{sec:virtual}Virtual corrections} 

Using on-shell relations, the virtual corrections to the amplitude $M_0$ can always be expressed in terms of a shift to $f_\pm$, i.e.
\begin{equation}
M_0\to M_0+\delta M_\mathrm{vir},\quad {\rm with}\quad f_\pm (t)\rightarrow f_\pm(t)+\delta f_\pm(y,z)~.
\end{equation}
The corrections $\delta f_\pm$ are generally complex functions of two variables $y,z$ instead of a single variable $t$. 
The corresponding change in the decay rate is:
\begin{eqnarray}
(\delta\Gamma_{K_{\ell 3}})_\mathrm{vir}&=&\frac{M_K}{256\pi^3}\int_{\mathcal{D}_3}dydz\:2G_F^2|V_{us}|^2\left\{H(+1,+1)f_+\mathfrak{Re}\delta f_+\right.\nonumber\\
&&\left.+H(+1,-1)[f_+\mathfrak{Re}\delta f_-+f_-\mathfrak{Re}\delta f_+]+H(-1,-1)f_-\mathfrak{Re}\delta f_-\right\}.
\end{eqnarray}
An important feature of the equation above is that both $H(+1,-1)$ and $H(-1,-1)$ contain an explicit factor $r_\ell\equiv m_\ell^2/M_K^2$ which suppresses the contribution of $\delta f_-$ to the decay rate. In the $K_{e3}$ channels, $r_e\sim 10^{-6}$ and $\delta f_-$ is completely negligible, which is one of the basic assumptions of our previous work. However, even in the $K_{\mu 3}$ channels, the suppression $r_\mu\sim 0.04$ is still quite significant. Therefore, apart from the IR-singular pieces in $\delta f_-$ that need to be rigorously calculated for the exact cancellation of IR-divergences, one is allowed to adopt an approximate representation for the remaining, IR-finite pieces without introducing a large error budget to the decay rate. This is the main spirit of this work.

The general structure of the virtual corrections to the decay amplitude can be summarized following Section 3 of Ref.~\cite{Seng:2021wcf} as
\begin{align}
\delta M_\mathrm{vir}=\delta M_\mathrm{I}+(\delta M_2+\delta M_{\gamma W}^a)_\mathrm{int}+\delta M_{\gamma W}^{b}+\delta M_3.
\end{align}
The first term collects all universal, model-independent analytical pieces that originate from corrections sensitive to IR and UV physics.
The second and third terms combine the remnants of the two-point function and the $\gamma W$-box contributions upon accounting for the respective analytic contributions in the first term. The last term represents the contribution of the three-point function. As explained in Ref.~\cite{Seng:2021wcf}, the full $\gamma W$-box correction stems from the generalized Compton tensor $T^{\mu\nu}$ which contains a symmetric and an antisymmetric pieces. 
Correspondingly, the two contributions $\delta M_{\gamma W}^a$ and $\delta M_{\gamma W}^b$ are distinguished. 
Below, we discuss all different contributions to $\delta f_\pm$ in Sirlin's representation in necessary detail.

\subsection{Analytic pieces}

Firstly, part of the loop integrals in the $\mathcal{O}(\alpha)$ electroweak RC can be calculated analytically (bearing errors of the order $\mathcal{O}(m_\ell^2/M_W^2)$) and the outcome is proportional to the tree-level amplitude, 
\begin{equation}
(\delta f_\pm)_\mathrm{I}=\left\{\frac{\alpha}{2\pi}\left[\ln\frac{M_Z^2}{m_\ell^2}-\frac{1}{4}\ln\frac{M_W^2}{m_\ell^2}+\frac{1}{2}\ln\frac{m_\ell^2}{M_\gamma^2}-\frac{3}{8}+\frac{1}{2}\tilde{a}_g\right]+\frac{1}{2}\delta_\mathrm{HO}^\mathrm{QED}\right\}f_\pm (t)~,
\end{equation}
see Eq.~(3.11) and below of Ref.~\cite{Seng:2021wcf} for explanations of the notations. Throughout this work we introduce a fictitious photon mass $M_{\gamma}$ to regularize the IR-divergence.

\subsection{\label{sec:deltaMint}From $(\delta M_2+\delta M_{\gamma W}^a)_\mathrm{int}+\delta M_{\gamma W}^{b,V}$}

Next we have the integrals $(\delta M_2+\delta M_{\gamma W}^a)_\mathrm{int}+\delta M_{\gamma W}^{b,V}$ defined in 
Eqs.~(3.9) and (3.10) of Ref.~\cite{Seng:2021wcf}. Two relevant comments are in order:
\begin{itemize}
	\item They give IR-divergent contributions to both $\delta f_\pm$.
	\item One can show that their contributions to $\delta f_+$ cannot depend on physics at large loop momentum; 
	the same argument, however, does not apply to $\delta f_-$.
\end{itemize}
Therefore, we adopt the following strategy:
\begin{enumerate}
	\item To study $\delta f_+$ from these integrals, we apply the same technique as in Sec.~4 of Ref.~\cite{Seng:2021wcf}, to write
	\begin{equation}
	(\delta f_+)_\mathrm{this\:subsection}=(\delta f_+)_\mathrm{II}+(\delta f_+)_\mathrm{conv}^\mathrm{fin}+(\delta f_+)_\mathrm{Born-conv}+(\delta f_+)_\mathrm{inel}~.\label{eq:deltafpint}
	\end{equation}
	 The calculable piece is the so-called ``Born contribution'' to the integral, and within it is the ``convection term'' contribution that is analytically calculable and contains the full IR-divergence. We split the latter into the IR divergent piece $(\delta f_+)_\mathrm{II}$ and the IR-finite piece $(\delta f_+)_\mathrm{conv}^\mathrm{fin}$. The remaining ``Born minus convection'' contribution can be computed numerically making use of the $\pi^-$ and $K^+$ electromagnetic form factors, for which we adopt a monopole parameterization~\cite{Amendolia:1986wj,Amendolia:1986ui}. Finally, the systematic uncertainty due to the incalculable inelastic (or ``non-Born'') pieces is estimated using the resonance chiral theory~\cite{Ecker:1988te,Ecker:1989yg,Cirigliano:2006hb}, see Appendix~B of Ref.~\cite{Seng:2021wcf} for details. We find that their corrections to $\delta_{K_{\ell 3}}$ in all channels are at most $1\times 10^{-4}$, so we simply assign a conservative uncertainty of $2\times 10^{-4}$ to them in each channel.
	 \item For $\delta f_-$, we do not attempt to calculate the full Born contribution because the convergence of the loop integrals depends on the ultraviolet (UV)-behavior of the form factors, which introduces extra model-dependence to the outcome. Instead, we compute only the convection term contribution (both the IR-divergent and IR-finite pieces, of which loop integrals are explicitly UV-finite) and leave the remaining, non-convection contribution as an incalculable piece at the moment:
	 \begin{equation}
	 (\delta f_-)_\mathrm{this\:subsection}=(\delta f_-)_\mathrm{II}+(\delta f_-)_\mathrm{conv}^\mathrm{fin}+(\delta f_-)_\mathrm{non-conv}~.
	 \end{equation} 
\end{enumerate}

Below we provide the analytic expressions of the convection term contribution to $\delta f_\pm$ from $(\delta M_2+\delta M_{\gamma W}^a)_\mathrm{int}+\delta M_{\gamma W}^{b,V}$:
\begin{eqnarray}
(\delta f_{\pm}^{K^0\pi^-})_\mathrm{II}&=&-\frac{\alpha}{4\pi}\left\{-\frac{4p'\cdot p_\ell x_s}{M_\pi m_\ell (1-x_s^2)}\ln x_s\ln\left(\frac{M_\gamma^2}{M_\pi m_\ell}\right)f_{\pm}^{K^0\pi^-}\right.\nonumber\\
&&\left.\pm\left(\frac{5}{2}-\ln\frac{M_\pi^2}{M_\gamma^2}\right)\left(\frac{p'\cdot(p+p')}{2M_\pi^2}f_+^{K^0\pi^-}+\frac{p'\cdot(p-p')}{2M_\pi^2}f_-^{K^0\pi^-}\right)\right\}\nonumber\\
(\delta f_{\pm}^{K^+\pi^0})_\mathrm{II}&=&-\frac{\alpha}{4\pi}\left\{\frac{4p\cdot p_\ell x_u}{M_K m_\ell (1-x_u^2)}\ln x_u\ln\left(\frac{M_\gamma^2}{M_K m_\ell}\right)f_{\pm}^{K^+\pi^0}\right.\nonumber\\
&&\left.+\left(\frac{5}{2}-\ln\frac{M_K^2}{M_\gamma^2}\right)\left(\frac{p\cdot(p+p')}{2M_K^2}f_+^{K^+\pi^0}+\frac{p\cdot(p-p')}{2M_K^2}f_-^{K^+\pi^0}\right)\right\}
\end{eqnarray}

\begin{eqnarray}
(\delta f_+^{K^0\pi^-})_\mathrm{conv}^\mathrm{fin}&=&-\frac{\alpha}{4\pi}\left\{\left(C_{00}^\mathrm{fin}+4p'\cdot p_\ell C_0^\mathrm{fin}+2p'\cdot p_\ell C_1-2m_\ell^2 C_2\right)f_+^{K^0\pi^-}\right.\nonumber\\
&&+\left(p'\cdot(p+p')f_+^{K^0\pi^-}+p'\cdot(p-p')f_-^{K^0\pi^-}\right)(C_1+C_{11}/2)\nonumber\\
&&-\left(p_\ell\cdot(p+p')f_+^{K^0\pi^-}+p_\ell\cdot(p-p')f_-^{K^0\pi^-}\right)C_{12}/2\nonumber\\
&&\left.+\left(p_\ell\cdot(p'-p)+m_\ell^2\right)\left(f_+^{K^0\pi^-}+f_-^{K^0\pi^-}\right)C_2\right\}\nonumber\\
(\delta f_-^{K^0\pi^-})_\mathrm{conv}^\mathrm{fin}&=&-\frac{\alpha}{4\pi}\left\{\left(C_{00}^\mathrm{fin}+4p'\cdot p_\ell C_0^\mathrm{fin}+2p'\cdot p_\ell C_1-2m_\ell^2 C_2\right)f_-^{K^0\pi^-}\right.\nonumber\\
&&-\left(p'\cdot(p+p')f_+^{K^0\pi^-}+p'\cdot(p-p')f_-^{K^0\pi^-}\right)(C_1+C_{11}/2+2C_2+C_{12})\nonumber\\
&&+\left(p_\ell\cdot(p+p')f_+^{K^0\pi^-}+p_\ell\cdot(p-p')f_-^{K^0\pi^-}\right)(C_{12}/2+C_{22})\nonumber\\
&&\left.+\left(p_\ell\cdot(p-3p')+2p\cdot p'-2M_\pi^2-m_\ell^2\right)\left(f_+^{K^0\pi^-}+f_-^{K^0\pi^-}\right)C_2\right\}
\end{eqnarray}

\begin{eqnarray}
(\delta f_+^{K^+\pi^0})_\mathrm{conv}^\mathrm{fin}&=&-\frac{\alpha}{4\pi}\left\{\left(C_{00}^\mathrm{fin}-4p\cdot p_\ell C_0^\mathrm{fin}-2p\cdot p_\ell C_1-2m_\ell^2 C_2\right)f_+^{K^+\pi^0}\right.\nonumber\\
&&+\left(p\cdot(p+p')f_+^{K^+\pi^0}+p\cdot(p-p')f_-^{K^+\pi^0}\right)(C_1+C_{11}/2)\nonumber\\
&&+\left(p_\ell\cdot(p+p')f_+^{K^+\pi^0}+p_\ell\cdot(p-p')f_-^{K^+\pi^0}\right)C_{12}/2\nonumber\\
&&\left.+\left(p_\ell\cdot(p'-p)+m_\ell^2\right)\left(f_+^{K^+\pi^0}-f_-^{K^+\pi^0}\right)C_2\right\}\nonumber\\
(\delta f_-^{K^+\pi^0})_\mathrm{conv}^\mathrm{fin}&=&-\frac{\alpha}{4\pi}\left\{\left(C_{00}^\mathrm{fin}-4p\cdot p_\ell C_0^\mathrm{fin}-2p\cdot p_\ell C_1-2m_\ell^2 C_2\right)f_-^{K^+\pi^0}\right.\nonumber\\
&&+\left(p\cdot(p+p')f_+^{K^+\pi^0}+p\cdot(p-p')f_-^{K^+\pi^0}\right)(C_1+C_{11}/2+2C_2+C_{12})\nonumber\\
&&+\left(p_\ell\cdot(p+p')f_+^{K^+\pi^0}+p_\ell\cdot(p-p')f_-^{K^+\pi^0}\right)(C_{12}/2+C_{22})\nonumber\\
&&\left.+\left(p_\ell\cdot(p-3p')+2p\cdot p'-2M_\pi^2-m_\ell^2\right)\left(f_+^{K^+\pi^0}-f_-^{K^+\pi^0}\right)C_2\right\}
\end{eqnarray}
We refer the reader to Appendix~C in Ref.~\cite{Seng:2021wcf} for the definitions of $x_v$ and the $C$-functions; in $K_{\ell 3}^0$ their arguments are $m_1=M_\pi$, $m_2=m_\ell$, $v=s$, whereas in $K_{\ell 3}^+$ their arguments are $m_1=M_K$, $m_2=m_\ell$, $v=u$. We defer the discussion of their numerical impact on $\delta_{K_{\ell 3}}$ (as well as the ``Born minus convection'' contribution to $\delta f_+$) to Sections~\ref{sec:real} and \ref{sec:num}.

\subsection{From the three-point function}

Next we have the so-called ``three-point function'' correction to the form factors which we denote as $\delta f_{\pm,3}$, see 
Eq.~(5.14) in Ref.~\cite{Seng:2019lxf}. The same reference provides an approximate expression for these quantities to $\mathcal{O}(e^2p^2)$ in the chiral power counting (see Sec.~8 of that paper), but na\"{\i}vely applying those expressions here will cause an incomplete cancellation of IR-divergences. Therefore, adopting the strategy in Sec.~5 of Ref.~\cite{Seng:2021wcf}, we split the full $\delta f_{\pm,3}$ into the IR-divergent and the IR-finite pieces:
\begin{equation}
\delta f_{\pm,3}=(\delta f_\pm)_\mathrm{III}+(\delta f_{\pm,3})_\mathrm{fin}~.
\end{equation} 
Starting from its $\mathcal{O}(e^2p^2)$ expression, one could resum the IR-divergent piece $(\delta f_\pm)_\mathrm{III}$ to all orders in the chiral expansion by simply adding back the full form factors and to ensure the exact cancellation of IR-divergences between the real and virtual corrections, as we pointed out in Sec.~5 of Ref.~\cite{Seng:2021wcf}. The resummed version reads: 
\begin{eqnarray}
(\delta f_\pm^{K^0\pi^-})_\mathrm{III}&=&\mp\frac{\alpha}{4\pi}\left(\ln\frac{M_\pi^2}{M_\gamma^2}-\frac{5}{2}\right)\frac{p'\cdot (p\mp p')}{2M_\pi^2}\left(f_+^{K^0\pi^-}+f_-^{K^0\pi^-}\right)\nonumber\\
(\delta f_\pm^{K^+\pi^0})_\mathrm{III}&=&\pm\frac{\alpha}{4\pi}\left(\ln\frac{M_K^2}{M_\gamma^2}-\frac{5}{2}\right)\frac{p\cdot(p\mp p')}{2M_K^2}\left(f_+^{K^+\pi^0}-f_-^{K^+\pi^0}\right)~.
\end{eqnarray}

In the following, it is sufficient to adopt an $\mathcal{O}(e^2p^2)$ approximation to $(\delta f_{\pm,3})_\mathrm{fin}$, which could be inferred from Ref.~\cite{Seng:2019lxf}. However, there is one further complication as discussed in Sec.~8 of 
Ref.~\cite{Seng:2021wcf}, namely: Following the standard ChPT treatment, some of the terms in $(\delta f_{\pm,3})_\mathrm{fin}$ are in fact not counted as a part of the long-distance electromagnetic corrections, but rather redistributed into the extra $t$-dependence of the form factors (Eqs.~(5.3), (5.7), (5.11), (5.12) in Ref.~\cite{Cirigliano:2001mk}) as well as the isospin-breaking correction $\delta_\mathrm{SU(2)}^{K\pi}$ through the electromagnetically-induced $\pi^0-\eta$ mixing (Eq.~(5.5) in 
Ref.~\cite{Cirigliano:2001mk}). Therefore, in order to appropriately compare with the  existing literature, we have to remove these terms from our definitions of $(\delta f_{\pm,3})_\mathrm{fin}$ as well. After doing so, we obtain the following $\mathcal{O}(e^2p^2)$ expressions:
\begin{eqnarray}
(\delta f_{+,3}^{K^0\pi^-})_\mathrm{fin}^{e^2p^2}&=&0\nonumber\\
(\delta f_{+,3}^{K^+\pi^0})_\mathrm{fin}^{e^2p^2}&=&0\nonumber\\
(\delta f_{-,3}^{K^0\pi^-})_\mathrm{fin}^{e^2p^2}&=&\frac{\alpha}{4\pi}\left[\frac{3}{4}\ln\frac{M_\pi^2}{\mu^2}-1\right]-\frac{Z\alpha}{8\pi}\ln\frac{M_\pi^2}{\mu^2}+\frac{8\pi\alpha}{3}\left(K_5^r+K_6^r\right)\nonumber\\
(\delta f_{-,3}^{K^+\pi^0})_\mathrm{fin}^{e^2p^2}&=&-\frac{\alpha}{4\sqrt{2}\pi}\left[\frac{3}{4}\ln\frac{M_K^2}{\mu^2}-1\right]-\frac{5Z\alpha}{8\sqrt{2}\pi}\ln\frac{M_K^2}{\mu^2}\nonumber\\
&&+\frac{8\pi\alpha}{\sqrt{2}}\left(-2K_3^r+K_4^r+\frac{1}{3}K_5^r+\frac{1}{3}K_6^r\right)~,
\end{eqnarray}
where $Z\approx 0.8$ and $K_i^r$ are low-energy constants (LECs) in ChPT, with $\mu$ the scale of dimensional regularization, which will be taken as $M_\rho=770$~MeV in the numerical analysis. 

\subsection{From $\delta M_{\gamma W}^{b,A}$}

Finally, we have the contribution from the axial $\gamma W$-box amplitude $\delta M_{\gamma W}^{b,A}$, namely the axial charged weak current contribution to the loop integral $\delta M_{\gamma W}^b$ defined in Eq.~(3.10) of 
Ref.~\cite{Seng:2021wcf}. In Sec.~6 of that paper, we outlined the strategy to fix its contribution to $\delta f_+$, namely to make use of the existing lattice QCD calculations of the forward axial $\gamma W$-box diagrams in the meson sector. In short, we can write:
\begin{equation}
(\delta f_+)_{\gamma W}^{b,A}=\left\{\Box_{\gamma W}^{VA>}+\left[\Box_{\gamma W}^{VA<}(K,\pi,M_\pi)+\mathcal{O}\left(\frac{M_K^2}{\Lambda_\chi^2}\right)\right]\right\}f_+(t)~.
\end{equation}
First, $\Box_{\gamma W}^{VA>}\approx 2.16\times 10^{-3}$ is the channel-independent component originating from the integral of the forward box diagram at $Q^2>2$~GeV$^2$, which was calculated to high precision from perturbative QCD. In the meantime, $\Box_{\gamma W}^{VA<}(K,\pi,M_\pi)$ is the remaining, channel-dependent part of the forward $K\pi$ box diagram calculated in the flavor SU(3) limit $M_K=M_\pi$. Existing lattice calculations are for $\Box_{\gamma W}^{VA<}(\pi^+,\pi^0,M_\pi)$~\cite{Feng:2020zdc} and $\Box_{\gamma W}^{VA<}(K^0,\pi^-,M_\pi)$~\cite{Ma:2021azh}, from which we can also obtain $\Box_{\gamma W}^{VA<}(K^+,\pi^0,M_\pi)=2\Box_{\gamma W}^{VA<}(\pi^+,\pi^0,M_\pi)-\Box_{\gamma W}^{VA<}(K^0,\pi^-,M_\pi)$ through a ChPT matching~\cite{Seng:2020jtz}. The uncertainties due to the non-forward (NF) corrections are conservatively estimated by multiplying the central values of $\Box_{\gamma W}^{VA<}$ by the factor $M_K^2/\Lambda_\chi^2$, where $\Lambda=4\pi F_\pi\approx 1.2$~GeV is the chiral symmetry breaking scale.

Unfortunately, these lattice calculations of the forward axial $\gamma W$-box diagrams do not provide any useful information for $(\delta f_-)_{\gamma W}^{b,A}$ because $p-p'=0$ by definition. So at this point we just leave it as another incalculable piece. 

\subsection{Matching the incalculable pieces in $\delta f_-$ to ChPT}

So far we have discussed the full $\mathcal{O}(\alpha)$ electroweak RC to $\delta f_\pm$. For $\delta f_+$ we have:
\begin{equation}
\delta f_+=(\delta f_+)_\mathrm{I+II+III}+(\delta f_+)_\mathrm{conv}^\mathrm{fin}+(\delta f_+)_\mathrm{Born-conv}+(\delta f_+)_\mathrm{inel}+(\delta f_{+,3})_\mathrm{fin}+(\delta f_+)_{\gamma W}^{b,A}~.
\end{equation}
From the discussions above, we see that each term at the right-hand side can either be determined to high precision or is small enough to assign a controllable theory uncertainty. 

Meanwhile, for $\delta f_-$ we have:
\begin{equation}
\delta f_-=(\delta f_-)_\mathrm{I+II+III}+(\delta f_-)_{\mathrm{conv}}^\mathrm{fin}+(\delta f_-)_\mathrm{non-conv}+(\delta f_{-,3})_\mathrm{fin}+(\delta f_-)_{\gamma W}^{b,A}~,\label{eq:fmsplit}
\end{equation}
among which $(\delta f_-)_\mathrm{I+II+III}$ and $(\delta f_-)_\mathrm{conv}^\mathrm{fin}$ are exactly known, $(\delta f_{-,3})_\mathrm{fin}$ is known to $\mathcal{O}(e^2p^2)$, while $(\delta f_-)_\mathrm{non-conv}$ and $(\delta f_-)_{\gamma W}^{b,A}$ are so far completely unknown. Fortunately, since they are IR-regular terms whose contributions to $\delta_{K_{\ell 3}}$ are naturally suppressed by $r_\ell$, there is a simple strategy to deal with them: Existing ChPT studies provided the $\mathcal{O}(e^2p^2)$ expression for the full $\delta f_-$ (i.e. the left-hand side of Eq.~\eqref{eq:fmsplit})~\cite{Cirigliano:2001mk}, and we can further take the $\mathcal{O}(e^2p^2)$ approximation for $(\delta f_-)_\mathrm{I+II+III}$ and $(\delta f_-)_\mathrm{conv}^\mathrm{fin}$ by replacing:
\begin{equation}
f_+^{K^0\pi^-}(t)\rightarrow-1~,~f_+^{K^+\pi^0}(t)\rightarrow -\frac{1}{\sqrt{2}}~,~f_-(t)\rightarrow 0~.
\end{equation}  
By doing so, one could then equate both sides in Eq.~\eqref{eq:fmsplit} to extract the $\mathcal{O}(e^2p^2)$ expression of $(\delta f_-)_\mathrm{non-conv}+(\delta f_-)_{\gamma W}^{b,A}$. As a consequence, we arrive the following representation:
\begin{equation}
\delta f_-= (\delta f_-)_\mathrm{I+II+III}+(\delta f_-)_{\mathrm{conv}}^\mathrm{fin}+(\delta f_-)_\mathrm{rem}
\end{equation}
where the first two terms at the right-hand side are exactly known, while the remaining piece  $(\delta f_-)_\mathrm{rem}$ adopts an $\mathcal{O}(e^2p^2)$ approximation:
\begin{eqnarray}
(\delta f_-^{K^0\pi^-})_\mathrm{rem}^{e^2p^2}&=&-\frac{\alpha}{4\pi}\left[r_{0,1}\Lambda(s,M_\pi,m_\ell)+r_{0,2}\ln\frac{M_\pi^2}{m_\ell^2}+r_{0,3}\ln\frac{M_\pi^2}{\mu^2}+r_{0,4}\right]\nonumber\\
(\delta f_-^{K^+\pi^0})_\mathrm{rem}^{e^2p^2}&=&-\frac{\alpha}{4\sqrt{2}\pi}\left[r_{+,1}\Lambda(u,M_K,m_\ell)+r_{+,2}\ln\frac{M_K^2}{m_\ell^2}+r_{+,3}\ln\frac{M_K^2}{\mu^2}+r_{+,4}\right]~,
\end{eqnarray}
where the function $\Lambda(v,m_1,m_2)$ is defined in Eq.~(C.5) of Ref.~\cite{Seng:2021wcf}. The coefficients $r_{i,j}$ are given by:
\begin{eqnarray}
r_{0,1}&=&-\frac{1}{8\lambda s}\left\{M_K^2[2s(m_\ell^2-2M_\pi^2)+(m_\ell^2-M_\pi^2)^2+3s^2]-m_\ell^6+m_\ell^4[3M_\pi^2+s]\right.\nonumber\\
&&-m_\ell^2[3M_\pi^4+13M_\pi^2s+s(t-20s)]+M_\pi^6+12M_\pi^4s+7M_\pi^2s^2+st[M_\pi^2-5s]\nonumber\\
&&\left.-20s^3\right\}\nonumber\\
r_{0,2}&=&\frac{1}{16s^2}[M_K^2(-m_\ell^2+M_\pi^2-3s)+m_\ell^4-2m_\ell^2M_\pi^2+M_\pi^4+13M_\pi^2s+28s^2+st]\nonumber\\
r_{0,3}&=&\frac{Z-6}{2}\nonumber\\
r_{0,4}&=&32\pi^2[X_1+X_2^r-X_3^r-(K_5^r+K_6^r)/3]+\frac{m_\ell^2-M_K^2-M_\pi^2}{8s}+\frac{3}{4}
\end{eqnarray}
and
\begin{eqnarray}
r_{+,1}&=&\frac{1}{8\lambda u}\left\{u^2[7M_K^2+20m_\ell^2+3M_\pi^2-5t]+[M_K^2-m_\ell^2]^2[M_K^2-m_\ell^2+M_\pi^2]\right.\nonumber\\
&&\left.+u[12M_K^4+M_K^2(-13m_\ell^2-4M_\pi^2+t)+m_\ell^2(m_\ell^2+2M_\pi^2-t)]-20u^3\right\}\nonumber\\
r_{+,2}&=&-\frac{1}{16u^2}[(M_K^2-m_\ell^2)(M_K^2-m_\ell^2+M_\pi^2)+u(13M_K^2-3M_\pi^2+t)+28u^2]\nonumber\\
r_{+,3}&=&\frac{5Z+6}{2}\nonumber\\
r_{+,4}&=&32\pi^2[X_1-X_2^r+X_3^r+2K_3^r-K_4^r-(K_5^r+K_6^6)/3]+\frac{M_K^2+M_\pi^2-m_\ell^2}{8u}-\frac{3}{4}~,
\end{eqnarray}
where $\lambda=\lambda(m_1^2,m_2^2,v)$ is the K\"{a}ll\'{e}n function.

\section{\label{sec:real}Real corrections}

Our previous treatment of the $K_{e3}$ bremsstrahlung correction detailed in Sec.~7 of Ref.~\cite{Seng:2021wcf} can be directly applied to all $K_{\ell 3}$ channels without any further modification. During the bremsstrahlung, the real photon can be emitted either by the charged lepton or by the meson; in the latter case, the decay amplitude involves the generalized, non-forward Compton tensor $T_{\mu\nu}(q';p',p)$ (see Eq.~(3.5) in Ref.~\cite{Seng:2021wcf}). Our strategy is to split $T_{\mu\nu}$ into the convection term contribution and a remainder:
\begin{equation}
T^{\mu\nu}=T^{\mu\nu}_\mathrm{conv}+[T^{\mu\nu}-T^{\mu\nu}_\mathrm{conv}]~.
\end{equation}
In practice, we retain the full $T^{\mu\nu}_\mathrm{conv}$ but take the $\mathcal{O}(p^2)$ approximation for $T^{\mu\nu}-T^{\mu\nu}_\mathrm{conv}$; this simple implementation ensures exact electromagnetic gauge invariance. Consequently, we split the full tree-level $K\rightarrow \pi\ell^+\nu_\ell \gamma$ amplitudes into two pieces:
\begin{equation}
M_{K\rightarrow \pi\ell^+\nu_\ell\gamma}=M_A+M_B~,
\end{equation} 
where $T^{\mu\nu}-T^{\mu\nu}_\mathrm{conv}$ is fully contained in $M_B$. Therefore, in calculating the squared amplitude, the result of $|M_A|^2$ is exact while $2\mathfrak{Re}\{M_A^*M_B\}+|M_B|^2$ acquires a chiral expansion uncertainty. 

The bremsstrahlung process resides at a larger allowed phase space area $\mathcal{D}_4=\mathcal{D}_3\oplus\mathcal{D}_{4-3}$ (see Appendix~A of Ref.~\cite{Seng:2021wcf}). The IR-divergence is fully contained in $|M_A|^2$ integrating over the $\mathcal{D}_3$ region, and is exactly canceled by those in $(\delta f_\pm)_\mathrm{I+II+III}$ from the virtual corrections.

\section{\label{sec:num}Numerical results, errors and correlations}

\begin{table}
	\begin{centering}
		\begin{tabular}{|c|c|c|c|c|}
			\hline 
			$\delta_{K_{\ell 3}}$& $(\delta f_+ )_{\mathrm{conv}}^{\mathrm{fin}}$ & $(\delta f_+)_{\mathrm{Born-conv}}$
			& $(\delta f_+)_{\gamma W}^{b,A}$&(*) $(\delta f_-)_\mathrm{rem}$
			\tabularnewline
			\hline 
			\hline 
			$K_{e3}^{0}$ & $-$5.0 & 4.1 & $4.9(1)_\mathrm{lat}(1)_\mathrm{NF}$
			&0.0\tabularnewline
			\hline 
			$K_{e3}^{+}$ & 9.6 & 0.1
			& $6.4(1)_\mathrm{lat}(4)_\mathrm{NF}$&0.0\tabularnewline
			\hline 
			$K_{\mu3}^{0}$ & 18.6 & 0.4&$4.9(1)_\mathrm{lat}(1)_\mathrm{NF}$& $0.3(2)_\mathrm{LEC}$
			\tabularnewline
			\hline 
			$K_{\mu3}^{+}$ & $-$0.1 & 1.0 & $6.5(1)_\mathrm{lat}(4)_\mathrm{NF}$&$-1.1(2)_\mathrm{LEC}$\tabularnewline
			\hline 
		\end{tabular}
		\par\end{centering}
	\caption{\label{tab:numvir}Contribution to $\delta_{K_{\ell 3}}$ from $(\delta f_+)_\mathrm{conv}^\mathrm{fin}$, $(\delta f_+)_\mathrm{Born-conv}$, $(\delta f_{+})_{\gamma W}^{b,A}$ and $(\delta f_-)_\mathrm{rem}$, in units of $10^{-3}$.}
	
\end{table}

\begin{table}
	\begin{centering}
		\begin{tabular}{|c|c|c|}
			\hline 
			$(\delta_{K_{\ell3}})_{\mathrm{I+II+III+brem}(\mathcal{D}_{3})}$ & (*) From $2\mathfrak{Re}\{M_{A}^{*}M_{B}\}+|M_{B}|^{2}$ & Remainder\tabularnewline
			\hline 
			\hline 
			$K_{e3}^{0}$ & 1.0 & 24.1\tabularnewline
			\hline 
			$K_{e3}^{+}$ & $-$0.3 & 4.4\tabularnewline
			\hline 
			$K_{\mu3}^{0}$ & 0.6 & 13.4\tabularnewline
			\hline 
			$K_{\mu3}^{+}$ & $-$0.1 & 17.4\tabularnewline
			\hline 
		\end{tabular}
		\par\end{centering}
	\caption{\label{tab:numD3}Sum of the IR-divergent one-loop contributions I, II, III and the
		bremsstrahlung contribution in the $\mathcal{D}_{3}$ region, in units of $10^{-3}$.}
	
\end{table}

\begin{table}
	\begin{centering}
		\begin{tabular}{|c|c|c|}
			\hline 
			$(\delta_{K_{\ell3}})_{\mathrm{brem}(\mathcal{D}_{4-3})}$ & (*) From $2\mathfrak{Re}\{M_{A}^{*}M_{B}\}+|M_{B}|^{2}$ & From $|M_{A}|^{2}$\tabularnewline
			\hline 
			\hline 
			$K_{e3}^{0}$ & 0.2 & 5.6\tabularnewline
			\hline 
			$K_{e3}^{+}$ & $-$0.1 & 5.3\tabularnewline
			\hline 
			$K_{\mu3}^{0}$ & 0.1 & 0.2\tabularnewline
			\hline 
			$K_{\mu3}^{+}$ & 0.0 & 0.1\tabularnewline
			\hline 
		\end{tabular}
		\par\end{centering}
	\caption{\label{tab:numD4m3}The bremsstrahlung contribution in the $\mathcal{D}_{4-3}$ region,
		in units of $10^{-3}$.}
	
\end{table}

Now we are ready to present the numerical results of $\delta_{K_{\ell 3}}$ in all channels. The IR-finite contribution from the virtual corrections is summarized in Table~\ref{tab:numvir}, the combination of the IR-divergent parts in the virtual corrections and the bremsstrahlung corrections in the $\mathcal{D}_3$ region is given in Table~\ref{tab:numD3}, and the bremsstrahlung contribution in the $\mathcal{D}_{4-3}$ region is given in Table~\ref{tab:numD4m3}. Columns in these three tables with an asterisk (*) denote quantities that have taken an $\mathcal{O}(e^2p^2)$ approximation and are subject to chiral uncertainties at $\mathcal{O}(e^2p^4)$. 

For the  error analysis we include only theory uncertainties that  after a single rounding-up are of the size $1\times 10^{-4}$ or larger. After subtracting out $S_\mathrm{EW}-1$ and its associated short-distance error, there are five major sources of theory uncertainties:\footnote{The uncertainty due to the kaon charge radius in $K_{e3}^+$ channel is around $4.8\times 10^{-5}$. In Ref.~\cite{Seng:2021wcf}, we somewhat exaggerated this uncertainty by first rounding it up to $5\times 10^{-5}$ and then displaying it as $1\times 10^{-4}$. In this work we do not perform such two-time rounding up so this uncertainty is not taken into account anymore.}
\begin{enumerate}
	\item The lattice uncertainty in $(\delta f_+)_{\gamma W}^{b,A}$, 
	\item The LEC uncertainties in $(\delta f_-)_\mathrm{rem}$,
	\item The NF uncertainty in $(\delta f_+)_{\gamma W}^{b,A}$,
	\item The $\mathcal{O}(e^2p^4)$ chiral uncertainty, and
	\item The unknown $(\delta f_+)_\mathrm{inel}$ in Eq.\eqref{eq:deltafpint}.
\end{enumerate}
The first three are already displayed in Table~\ref{tab:numvir}, but the last two not yet. In any case, in order to discuss later the average of $|V_{us}|$ over different channels, it is not sufficient to know just the uncertainties in each separate channel, but their correlations must also be understood. To do so one needs to first identify the independent theory inputs that carry the uncertainties, and then the correlation matrix can be calculated using the formula outlined in the Appendix of 
Ref.~\cite{Seng:2021nar}. This can be rigorously done for the first two types, which we describe below:
\begin{itemize}
	\item For $(\delta f_+)_{\gamma W}^{b,A}$, the independent lattice QCD inputs are~\cite{Feng:2020zdc,Ma:2021azh}:
	\begin{equation}
	\Box_{\gamma W}^{VA<}(\pi^+,\pi^0,M_\pi)=0.671(28)_\mathrm{lat}\times 10^{-3}~,~\Box_{\gamma W}^{VA<}(K^0,\pi^-,M_\pi)=0.278(44)_\mathrm{lat}\times 10^{-3}~.
	\end{equation}
	\item For $(\delta f_-)_\mathrm{rem}$, the independent combinations of LECs are $X_1$, $C_1\equiv X_2^r-X_3^r$, $C_2\equiv 2K_3^r-K_4^r$ and $C_3\equiv K_5^r+K_6^r$. Among them, $X_1=-2.2(4)\times 10^{-3}$ was fixed to good precision with the recent lattice calculations~\cite{Ma:2021azh}, and its resulting uncertainty to $\delta_{K_{\ell 3}}$ is negligible. Similar calculations are not yet done for $C_{1-3}$, so we infer their values at $\mu=M_\rho$ from resonance models~\cite{DescotesGenon:2005pw,Ananthanarayan:2004qk,Bijnens:2014lea}, and assign a 100\% uncertainty to each of them:
	\begin{equation}
	C_1=-1.4(1.4)_\mathrm{LEC}\times 10^{-3}~,~C_2=4.0(4.0)_\mathrm{LEC}\times 10^{-3}~,~C_3=14.4(14.4)_\mathrm{LEC}\times 10^{-3}~.
	\end{equation}  
\end{itemize}
Meanwhile, the next three uncertainties are estimated as follows:
\begin{itemize}
	\item The NF uncertainty in $(\delta f_+)_{\gamma W}^{b,A}$ is estimated by multiplying $\Box_{\gamma W}^{VA<}$ in each channel by $M_K^2/\Lambda_\chi^2$;
	\item The $\mathcal{O}(e^2p^4)$ chiral uncertainty is obtained by first adding all the columns with asterisks in Table~\ref{tab:numvir}--\ref{tab:numD4m3}, and then multiply the sum by $M_K^2/\Lambda_\chi^2$;
	\item Finally, a conservative uncertainty of $2\times 10^{-4}$ is assign to each channel to account for the poorly-constrained contribution from $(\delta f_+)_\mathrm{inel}$ (see discussions in Sec.\ref{sec:deltaMint}). 
\end{itemize}
Unlike the first two, these three errors are deduced using na\"{\i}ve power counting and order-of-magnitude estimations, and it is difficult to identify independent sources of uncertainties within each type. In fact, we consider it as arbitrary to take these uncertainties to be uncorrelated as to assume any correlation. Therefore, we simply take them to be uncorrelated, following the same strategy adopted by some of us in Ref.~\cite{Seng:2021nar}.

\begin{table}
	\begin{centering}
		\begin{tabular}{|c|c|c|}
			\hline 
			& $\delta_{\mathrm{EM}}^{K\ell}$&ChPT\tabularnewline
			\hline 
			\hline 
			$K^0e$ & $11.6(2)_{\mathrm{inel}}(1)_{\mathrm{lat}}(1)_{\mathrm{NF}}(2)_{e^{2}p^{4}}$&$9.9(1.9)_{e^2p^4}(1.1)_\mathrm{LEC}$\tabularnewline
			\hline 
			$K^+e$ & $2.1(2)_{\mathrm{inel}}(1)_{\mathrm{lat}}(4)_{\mathrm{NF}}(1)_{e^{2}p^{4}}$&$1.0(1.9)_{e^2p^4}(1.6)_\mathrm{LEC}$\tabularnewline
			\hline 
			$K^0\mu$ & $15.4(2)_{\mathrm{inel}}(1)_{\mathrm{lat}}(1)_{\mathrm{NF}}(2)_{\mathrm{LEC}}(2)_{e^{2}p^{4}}$&$14.0(1.9)_{e^2p^4}(1.1)_\mathrm{LEC}$\tabularnewline
			\hline 
			$K^+\mu$ & $0.5(2)_{\mathrm{inel}}(1)_{\mathrm{lat}}(4)_{\mathrm{NF}}(2)_{\mathrm{LEC}}(2)_{e^{2}p^{4}}$&$0.2(1.9)_{e^2p^4}(1.6)_\mathrm{LEC}$\tabularnewline
			\hline 
		\end{tabular}
		\par\end{centering}
	\caption{\label{tab:deltaEM}Final result for $\delta_{\mathrm{EM}}^{K\ell}$, in units of $10^{-3}$. The ChPT result from Ref.\cite{Cirigliano:2008wn} is given in the last column for comparison. }
	
\end{table}

We present our final result of $\delta_\mathrm{EM}^{K\ell}$ in Table~\ref{tab:deltaEM}, with the correlation matrix given by:
\begin{equation}
\mathrm{Corr}(\delta_{\mathrm{EM}})=\left(\begin{array}{cccc}
1 & -0.050 & 0.069 & -0.043\\
& 1 & -0.049 & 0.079\\
&  & 1 & 0.088\\
&  &  & 1
\end{array}\right)~,
\end{equation}
where $\delta_{\mathrm{EM}}=\left(\begin{array}{cccc}
\delta_{\mathrm{EM}}^{K^{0}e} & \delta_{\mathrm{EM}}^{K^{+}e} & \delta_{\mathrm{EM}}^{K^{0}\mu} & \delta_{\mathrm{EM}}^{K^{+}\mu}\end{array}\right)^{T}$. 
The results in the two $Ke$ channels are obviously the same as in Ref.~\cite{Seng:2021wcf}. As a comparison, we also quote the ChPT result from Ref.~\cite{Cirigliano:2008wn}. Our new determinations agree with them within error bars in all four channels, but with significant improvements in precision by almost an order of magnitude.

\begin{table}
	\begin{centering}
		\begin{tabular}{|c|c|cccccc|}
			\hline 
			& $|V_{us}f_{+}^{K^0\pi^-}(0)|$ & \multicolumn{6}{c|}{Correlation Matrix}\tabularnewline
			\hline 
			\hline 
			$K_{L}e$ & $0.21617(46)_{\mathrm{exp}}(10)_{I_{K}}(4)_{\delta_{\mathrm{EM}}}\,\,\,\,$ & 1 & 0.021 & 0.025 & 0.567 & 0.004 & 0.017\tabularnewline
			\cline{1-2} 
			$K_{S}e$ & $0.21530(122)_{\mathrm{exp}}(10)_{I_{K}}(4)_{\delta_{\mathrm{EM}}}\,\,$ &  & 1 & 0.009 & 0.014 & 0.000 & 0.006\tabularnewline
			\cline{1-2} 
			$K^{+}e$ & $\,\,\,\,\,\,\,\,\,\,\,\,\,\,\,\,\,\,0.21714(88)_{\mathrm{exp}}(10)_{I_{K}}(21)_{\delta_{\mathrm{SU(2)}}}(5)_{\delta_{\mathrm{EM}}}$ &  &  & 1 & 0.018 & 0.002 & 0.894\tabularnewline
			\cline{1-2} 
			$K_{L}\mu$ & $0.21649(50)_{\mathrm{exp}}(16)_{I_{K}}(4)_{\delta_{\mathrm{EM}}}\,\,$ &  &  &  & 1 & 0.011 & 0.044\tabularnewline
			\cline{1-2} 
			$K_{S}\mu$ & $\,\,0.21251(466)_{\mathrm{exp}}(16)_{I_{K}}(4)_{\delta_{\mathrm{EM}}}\,$ &  &  &  &  & 1 & 0.005\tabularnewline
			\cline{1-2} 
			$K^{+}\mu$ & $\,\,\,\,\,\,\,\,\,\,\,\,\,\,\,\,\,\,\,\,\,\,\,\,\,0.21699(108)_{\mathrm{exp}}(16)_{I_{K}}(21)_{\delta_{\mathrm{SU(2)}}}(6)_{\delta_{\mathrm{EM}}}$ &  &  &  &  &  & 1\tabularnewline
			\hline 
			Average: $Ke$ & $0.21626(40)_K(3)_\mathrm{HO}$ &  &  &  &  &  & \multicolumn{1}{c}{}\tabularnewline
			\cline{1-2} 
			Average: $K\mu$ & $0.21654(48)_K(3)_\mathrm{HO}$ &  &  &  &  &  & \multicolumn{1}{c}{}\tabularnewline
			\cline{1-2} 
			Average: tot & $0.21634(38)_{K}(3)_{\mathrm{HO}}$ &  &  &  &  &  & \multicolumn{1}{c}{}\tabularnewline
			\cline{1-2} 
		\end{tabular}
		\par\end{centering}
	\caption{\label{tab:Vusf} Updated determination of $|V_{us}f_+^{K^0\pi^-}(0)|$, which should be compared to Table I in Ref.\cite{Seng:2021nar}.}
	
\end{table}

With these new theory inputs of $\delta_\mathrm{EM}^{K\ell}$, we also present here an updated global analysis of $|V_{us}f_+^{K^0\pi^-}(0)|$ from semileptonic kaon decays. Essentially, we take all the theory and experimental inputs quoted in Ref.~\cite{Seng:2021nar}, and only replace the values of $\delta_\mathrm{EM}$ and Corr($\delta_\mathrm{EM}$) by our new results. The outcome is summarized in Table~\ref{tab:Vusf}, which has only very minor changes comparing to Table~I in Ref.~\cite{Seng:2021nar}. Supplementing the result with the most recent FLAG averages of $|f_+^{K^0\pi^-}(0)|$ with different numbers of active quark flavors~\cite{Aoki:2021kgd}:
\begin{equation}
|f_+^{K^0\pi^-}(0)|=\left\{
\begin{array}{ccc}
0.9698(17) & N_f=2+1+1 & \quad\quad\text{Refs.~\cite{Carrasco:2016kpy,Bazavov:2018kjg}}\\
0.9677(27) & N_f=2+1 & \quad\quad\text{Refs.~\cite{Bazavov:2012cd,Boyle:2015hfa}}
\end{array}
\right.~,
\end{equation}
we obtain:
\begin{equation}
|V_{us}|_{K_{\ell 3}}=\left\{
\begin{array}{cc}
0.22308(39)_\mathrm{lat}(39)_K(3)_\mathrm{HO} & N_f=2+1+1 \\
0.22356(62)_\mathrm{lat}(39)_K(3)_\mathrm{HO} & N_f=2+1 
\end{array}
\right.\label{eq:VusKl3}
\end{equation}
respectively.

The value of $|V_{us}|$ can also be extracted by first obtaining $|V_{us}/V_{ud}|$ from the ratio between the leptonic kaon ($K_{\mu 2}$) and pion ($\pi_{\mu 2}$) decay rates, and then substituting $|V_{ud}|$ by its most precise determination from superallowed nuclear decays~\cite{Marciano:2004uf}. With this method, we obtain~\cite{Zyla:2020zbs}:
\begin{equation}
|V_{us}|_{K_{\mu 2}}=\left\{
\begin{array}{cc}
0.2252(5) & N_f=2+1+1 \\
0.2255(8) & N_f=2+1 
\end{array}
\right.~.\label{eq:VusKmu2}
\end{equation}
Comparing Eqs.~\eqref{eq:VusKl3} and \eqref{eq:VusKmu2}, we see that the so-called $K_{\ell 3}-K_{\mu 2}$ discrepancy in the $|V_{us}|$ determination is still very much alive.

\section{\label{sec:concl}Conclusion}

In this work we present for the first time a unified analysis of the long-distance electromagnetic RC to all channels in $K_{\ell 3}$ decays based on the novel  hybrid method combining Sirlin's representation with ChPT, that we developed a few years ago. With an appropriate use of the electromagnetic/charged weak form factors, lattice QCD inputs of the forward axial $\gamma W$-box as well as the existing knowledge of the remaining, poorly-constrained LECs, we are able to control the theory uncertainties in all four independent channels in $K_{\ell 3}$  at the level of $10^{-4}$. 

Several important implications of this new calculation are as follows. First, no significant change of  $\delta_\mathrm{EM}^{K\ell}$ from the previous ChPT result is observed; this allows us to state with more confidence that the outstanding discrepancy between the $|V_{us}|$ extracted from $K_{\mu 2}/\pi_{\mu 2}$ and $K_{\ell 3}$ does not originate from the long-distance electromagnetic corrections. Second, separate global analysis of $|V_{us}f_+^{K^0\pi^-}(0)|$ from $Ke$ and $K\mu$ channels show no significant difference, which supports the SM prediction of lepton flavor universality. Third, the further reduction of theory uncertainties provides an even stronger motivation for next-generation experiments to measure the kaon lifetimes and $K_{\ell 3}$ branching ratios with higher precision. From the theory side, the dominant sources of uncertainty are now the phase space factor $I_K$ and the isospin-breaking correction $\delta_\mathrm{SU(2)}$, which await new breakthroughs from phenomenology and lattice QCD.

Finally, we stress that every important theory input should always be cross-checked with several independent methods. In this sense, the existing plan to compute the full (virtual and real) $K_{\ell 3}$ RC directly on the  lattice~\cite{BoyleSnowmass} is very much anticipated.
	
\section*{Acknowledgements} 

This work is supported in
part by the Deutsche Forschungsgemeinschaft (DFG, German Research
Foundation) and the NSFC through the funds provided to the Sino-German Collaborative Research Center TRR110 ``Symmetries and the Emergence of Structure in QCD'' (DFG Project-ID 196253076 - TRR 110, NSFC Grant No. 12070131001) (U-G.M and C.Y.S), by the Chinese Academy of Sciences (CAS) through a President's
International Fellowship Initiative (PIFI) (Grant No. 2018DM0034) and by the VolkswagenStiftung
(Grant No. 93562) (U-G.M), by EU Horizon 2020 research and innovation programme, STRONG-2020 project
under grant agreement No 824093, by the German-Mexican research collaboration Grant No. 278017 (CONACyT)
and No. SP 778/4-1 (DFG), and partially by the DFG personal grant No GO 2604/3-1 (M.G).

	



\providecommand{\href}[2]{#2}\begingroup\raggedright\endgroup


\begin{thebibliography}{10}
	
	\bibitem{Sirlin:1977sv}
	A.~Sirlin, {\it {Current Algebra Formulation of Radiative Corrections in Gauge
			Theories and the Universality of the Weak Interactions}},  {\em Rev. Mod.
		Phys.} {\bf 50} (1978) 573. [Erratum: Rev. Mod. Phys.50,905(1978)].
	
	\bibitem{Seng:2021syx}
	C.-Y. Seng, {\it {Radiative Corrections to Semileptonic Beta Decays: Progress
			and Challenges}},  {\em Particles} {\bf 4} (2021), no.~4 397--467
	[\href{http://arXiv.org/abs/2108.03279}{{\tt 2108.03279}}].
	
	\bibitem{Seng:2018yzq}
	C.-Y. Seng, M.~Gorchtein, H.~H. Patel and M.~J. Ramsey-Musolf, {\it {Reduced
			Hadronic Uncertainty in the Determination of $V_{ud}$}},  {\em Phys. Rev.
		Lett.} {\bf 121} (2018), no.~24 241804
	[\href{http://arXiv.org/abs/1807.10197}{{\tt 1807.10197}}].
	
	\bibitem{Seng:2018qru}
	C.~Y. Seng, M.~Gorchtein and M.~J. Ramsey-Musolf, {\it {Dispersive evaluation
			of the inner radiative correction in neutron and nuclear $\beta$ decay}},
	{\em Phys. Rev.} {\bf D100} (2019), no.~1 013001
	[\href{http://arXiv.org/abs/1812.03352}{{\tt 1812.03352}}].
	
	\bibitem{Seng:2020wjq}
	C.-Y. Seng, X.~Feng, M.~Gorchtein and L.-C. Jin, {\it {Joint lattice
			QCD--dispersion theory analysis confirms the quark-mixing top-row unitarity
			deficit}},  {\em Phys. Rev. D} {\bf 101} (2020), no.~11 111301
	[\href{http://arXiv.org/abs/2003.11264}{{\tt 2003.11264}}].
	
	\bibitem{Shiells:2020fqp}
	K.~Shiells, P.~G. Blunden and W.~Melnitchouk, {\it {Electroweak axial structure
			functions and improved extraction of the Vud CKM matrix element}},  {\em
		Phys. Rev. D} {\bf 104} (2021), no.~3 033003
	[\href{http://arXiv.org/abs/2012.01580}{{\tt 2012.01580}}].
	
	\bibitem{Gorchtein:2021fce}
	M.~Gorchtein and C.-Y. Seng, {\it {Dispersion relation analysis of the
			radiative corrections to g$_{A}$ in the neutron \ensuremath{\beta}-decay}},
	{\em JHEP} {\bf 10} (2021) 053 [\href{http://arXiv.org/abs/2106.09185}{{\tt
			2106.09185}}].
	
	\bibitem{Gorchtein:2018fxl}
	M.~Gorchtein, {\it {$\gamma W$ Box Inside Out: Nuclear Polarizabilities Distort
			the Beta Decay Spectrum}},  {\em Phys. Rev. Lett.} {\bf 123} (2019), no.~4
	042503 [\href{http://arXiv.org/abs/1812.04229}{{\tt 1812.04229}}].
	
	\bibitem{Zyla:2020zbs}
	{\bf Particle Data Group} Collaboration, P.~Zyla {\em et.~al.}, {\it {Review of
			Particle Physics}},  {\em PTEP} {\bf 2020} (2020), no.~8 083C01.
	
	\bibitem{Fermigm2}
	{\bf Muon g-2} Collaboration, B.~Abi {\em et.~al.}, {\it {Measurement of the
			Positive Muon Anomalous Magnetic Moment to 0.46 ppm}},  {\em Phys. Rev.
		Lett.} {\bf 126} (2021) 141801.
	
	\bibitem{Aoyama:2020ynm}
	T.~Aoyama {\em et.~al.}, {\it {The anomalous magnetic moment of the muon in the
			Standard Model}},  {\em Phys. Rept.} {\bf 887} (2020) 1--166
	[\href{http://arXiv.org/abs/2006.04822}{{\tt 2006.04822}}].
	
	\bibitem{Miller:2007kk}
	J.~P. Miller, E.~de~Rafael and B.~L. Roberts, {\it {Muon (g-2): Experiment and
			theory}},  {\em Rept. Prog. Phys.} {\bf 70} (2007) 795
	[\href{http://arXiv.org/abs/hep-ph/0703049}{{\tt hep-ph/0703049}}].
	
	\bibitem{Miller:2012opa}
	J.~P. Miller, E.~de~Rafael, B.~L. Roberts and D.~St\"ockinger, {\it {Muon
			(g-2): Experiment and Theory}},  {\em Ann. Rev. Nucl. Part. Sci.} {\bf 62}
	(2012) 237--264.
	
	\bibitem{Jegerlehner:2009ry}
	F.~Jegerlehner and A.~Nyffeler, {\it {The Muon g-2}},  {\em Phys. Rept.} {\bf
		477} (2009) 1--110 [\href{http://arXiv.org/abs/0902.3360}{{\tt 0902.3360}}].
	
	\bibitem{Aaij:2019wad}
	{\bf LHCb} Collaboration, R.~Aaij {\em et.~al.}, {\it {Search for
			lepton-universality violation in $B^+\to K^+\ell^+\ell^-$ decays}},  {\em
		Phys. Rev. Lett.} {\bf 122} (2019), no.~19 191801
	[\href{http://arXiv.org/abs/1903.09252}{{\tt 1903.09252}}].
	
	\bibitem{Aaij:2014ora}
	{\bf LHCb} Collaboration, R.~Aaij {\em et.~al.}, {\it {Test of lepton
			universality using $B^{+}\rightarrow K^{+}\ell^{+}\ell^{-}$ decays}},  {\em
		Phys. Rev. Lett.} {\bf 113} (2014) 151601
	[\href{http://arXiv.org/abs/1406.6482}{{\tt 1406.6482}}].
	
	\bibitem{Aaij:2015yra}
	{\bf LHCb} Collaboration, R.~Aaij {\em et.~al.}, {\it {Measurement of the ratio
			of branching fractions $\mathcal{B}(\bar{B}^0 \to
			D^{*+}\tau^{-}\bar{\nu}_{\tau})/\mathcal{B}(\bar{B}^0 \to
			D^{*+}\mu^{-}\bar{\nu}_{\mu})$}},  {\em Phys. Rev. Lett.} {\bf 115} (2015),
	no.~11 111803 [\href{http://arXiv.org/abs/1506.08614}{{\tt 1506.08614}}].
	[Erratum: Phys.Rev.Lett. 115, 159901 (2015)].
	
	\bibitem{Aaij:2015oid}
	{\bf LHCb} Collaboration, R.~Aaij {\em et.~al.}, {\it {Angular analysis of the
			$B^{0} \to K^{*0} \mu^{+} \mu^{-}$ decay using 3 fb$^{-1}$ of integrated
			luminosity}},  {\em JHEP} {\bf 02} (2016) 104
	[\href{http://arXiv.org/abs/1512.04442}{{\tt 1512.04442}}].
	
	\bibitem{Seng:2019lxf}
	C.-Y. Seng, D.~Galviz and U.-G. Mei\ss{}ner, {\it {A New Theory Framework for
			the Electroweak Radiative Corrections in $K_{l3}$ Decays}},  {\em JHEP} {\bf
		02} (2020) 069 [\href{http://arXiv.org/abs/1910.13208}{{\tt 1910.13208}}].
	
	\bibitem{Seng:2020jtz}
	C.-Y. Seng, X.~Feng, M.~Gorchtein, L.-C. Jin and U.-G. Mei\ss{}ner, {\it {New
			method for calculating electromagnetic effects in semileptonic beta-decays of
			mesons}},  {\em JHEP} {\bf 10} (2020) 179
	[\href{http://arXiv.org/abs/2009.00459}{{\tt 2009.00459}}].
	
	\bibitem{Ma:2021azh}
	P.-X. Ma, X.~Feng, M.~Gorchtein, L.-C. Jin and C.-Y. Seng, {\it {Lattice QCD
			calculation of the electroweak box diagrams for the kaon semileptonic
			decays}},  {\em Phys. Rev. D} {\bf 103} (2021) 114503
	[\href{http://arXiv.org/abs/2102.12048}{{\tt 2102.12048}}].
	
	\bibitem{Cirigliano:2008wn}
	V.~Cirigliano, M.~Giannotti and H.~Neufeld, {\it {Electromagnetic effects in
			K(l3) decays}},  {\em JHEP} {\bf 11} (2008) 006
	[\href{http://arXiv.org/abs/0807.4507}{{\tt 0807.4507}}].
	
	\bibitem{Seng:2021boy}
	C.-Y. Seng, D.~Galviz, M.~Gorchtein and U.~G. Mei\ss{}ner, {\it {High-precision
			determination of the Ke3 radiative corrections}},  {\em Phys. Lett. B} {\bf
		820} (2021) 136522 [\href{http://arXiv.org/abs/2103.00975}{{\tt
			2103.00975}}].
	
	\bibitem{Seng:2021wcf}
	C.-Y. Seng, D.~Galviz, M.~Gorchtein and U.-G. Mei\ss{}ner, {\it {Improved
			$K_{e3}$ radiative corrections sharpen the $K_{\mu 2}$\textendash{}K$_{l3}$
			discrepancy}},  {\em JHEP} {\bf 11} (2021) 172
	[\href{http://arXiv.org/abs/2103.04843}{{\tt 2103.04843}}].
	
	\bibitem{Seng:2021nar}
	C.-Y. Seng, D.~Galviz, W.~J. Marciano and U.-G. Mei\ss{}ner, {\it {Update on
			$|V_{us}|$ and $|V_{us}/V_{ud}|$ from semileptonic kaon and pion decays}},
	{\em Phys. Rev. D} {\bf 105} (2022), no.~1 013005
	[\href{http://arXiv.org/abs/2107.14708}{{\tt 2107.14708}}].
	
	\bibitem{Marciano:1993sh}
	W.~J. Marciano and A.~Sirlin, {\it {Radiative corrections to pi(lepton 2)
			decays}},  {\em Phys. Rev. Lett.} {\bf 71} (1993) 3629--3632.
	
	\bibitem{Amendolia:1986wj}
	{\bf NA7} Collaboration, S.~Amendolia {\em et.~al.}, {\it {A Measurement of the
			Space - Like Pion Electromagnetic Form-Factor}},  {\em Nucl. Phys. B} {\bf
		277} (1986) 168.
	
	\bibitem{Amendolia:1986ui}
	S.~Amendolia {\em et.~al.}, {\it {A Measurement of the Kaon Charge Radius}},
	{\em Phys. Lett. B} {\bf 178} (1986) 435--440.
	
	\bibitem{Ecker:1988te}
	G.~Ecker, J.~Gasser, A.~Pich and E.~de~Rafael, {\it {The Role of Resonances in
			Chiral Perturbation Theory}},  {\em Nucl. Phys. B} {\bf 321} (1989) 311--342.
	
	\bibitem{Ecker:1989yg}
	G.~Ecker, J.~Gasser, H.~Leutwyler, A.~Pich and E.~de~Rafael, {\it {Chiral
			Lagrangians for Massive Spin 1 Fields}},  {\em Phys. Lett. B} {\bf 223}
	(1989) 425--432.
	
	\bibitem{Cirigliano:2006hb}
	V.~Cirigliano, G.~Ecker, M.~Eidemuller, R.~Kaiser, A.~Pich and J.~Portoles,
	{\it {Towards a consistent estimate of the chiral low-energy constants}},
	{\em Nucl. Phys. B} {\bf 753} (2006) 139--177
	[\href{http://arXiv.org/abs/hep-ph/0603205}{{\tt hep-ph/0603205}}].
	
	\bibitem{Cirigliano:2001mk}
	V.~Cirigliano, M.~Knecht, H.~Neufeld, H.~Rupertsberger and P.~Talavera, {\it
		{Radiative corrections to K(l3) decays}},  {\em Eur. Phys. J.} {\bf C23}
	(2002) 121--133 [\href{http://arXiv.org/abs/hep-ph/0110153}{{\tt
			hep-ph/0110153}}].
	
	\bibitem{Feng:2020zdc}
	X.~Feng, M.~Gorchtein, L.-C. Jin, P.-X. Ma and C.-Y. Seng, {\it
		{First-principles calculation of electroweak box diagrams from lattice QCD}},
	{\em Phys. Rev. Lett.} {\bf 124} (2020), no.~19 192002
	[\href{http://arXiv.org/abs/2003.09798}{{\tt 2003.09798}}].
	
	\bibitem{DescotesGenon:2005pw}
	S.~Descotes-Genon and B.~Moussallam, {\it {Radiative corrections in weak
			semi-leptonic processes at low energy: A Two-step matching determination}},
	{\em Eur. Phys. J.} {\bf C42} (2005) 403--417
	[\href{http://arXiv.org/abs/hep-ph/0505077}{{\tt hep-ph/0505077}}].
	
	\bibitem{Ananthanarayan:2004qk}
	B.~Ananthanarayan and B.~Moussallam, {\it {Four-point correlator constraints on
			electromagnetic chiral parameters and resonance effective Lagrangians}},
	{\em JHEP} {\bf 06} (2004) 047
	[\href{http://arXiv.org/abs/hep-ph/0405206}{{\tt hep-ph/0405206}}].
	
	\bibitem{Bijnens:2014lea}
	J.~Bijnens and G.~Ecker, {\it {Mesonic low-energy constants}},  {\em Ann. Rev.
		Nucl. Part. Sci.} {\bf 64} (2014) 149--174
	[\href{http://arXiv.org/abs/1405.6488}{{\tt 1405.6488}}].
	
	\bibitem{Aoki:2021kgd}
	Y.~Aoki {\em et.~al.}, {\it {FLAG Review 2021}},
	\href{http://arXiv.org/abs/2111.09849}{{\tt 2111.09849}}.
	
	\bibitem{Carrasco:2016kpy}
	N.~Carrasco, P.~Lami, V.~Lubicz, L.~Riggio, S.~Simula and C.~Tarantino, {\it
		{$K \to \pi$ semileptonic form factors with $N_f=2+1+1$ twisted mass
			fermions}},  {\em Phys. Rev. D} {\bf 93} (2016), no.~11 114512
	[\href{http://arXiv.org/abs/1602.04113}{{\tt 1602.04113}}].
	
	\bibitem{Bazavov:2018kjg}
	{\bf Fermilab Lattice, MILC} Collaboration, A.~Bazavov {\em et.~al.}, {\it
		{$|V_{us}|$ from $K_{\ell 3}$ decay and four-flavor lattice QCD}},  {\em
		Phys. Rev.} {\bf D99} (2019), no.~11 114509
	[\href{http://arXiv.org/abs/1809.02827}{{\tt 1809.02827}}].
	
	\bibitem{Bazavov:2012cd}
	A.~Bazavov {\em et.~al.}, {\it {Kaon semileptonic vector form factor and
			determination of $|V_{us}|$ using staggered fermions}},  {\em Phys. Rev. D}
	{\bf 87} (2013) 073012 [\href{http://arXiv.org/abs/1212.4993}{{\tt
			1212.4993}}].
	
	\bibitem{Boyle:2015hfa}
	{\bf RBC/UKQCD} Collaboration, P.~A. Boyle {\em et.~al.}, {\it {The kaon
			semileptonic form factor in N$_{f}$ = 2 + 1 domain wall lattice QCD with
			physical light quark masses}},  {\em JHEP} {\bf 06} (2015) 164
	[\href{http://arXiv.org/abs/1504.01692}{{\tt 1504.01692}}].
	
	\bibitem{Marciano:2004uf}
	W.~J. Marciano, {\it {Precise determination of |V(us)| from lattice
			calculations of pseudoscalar decay constants}},  {\em Phys. Rev. Lett.} {\bf
		93} (2004) 231803 [\href{http://arXiv.org/abs/hep-ph/0402299}{{\tt
			hep-ph/0402299}}].
	
	\bibitem{BoyleSnowmass}
	P.~Boyle {\em et.~al.}
	\newblock \textit{High-precision determination of $V_{us}$ and $V_{ud}$ from
		lattice QCD}.
	[\href{https://www.snowmass21.org/docs/files/summaries/RF/SNOWMASS21-RF2_RF0-TF5_TF0-CompF2_CompF0-054.pdf}{{\tt
			Link}}].
	
\end{thebibliography}
\end{document}